\documentclass[reprint,amsmath,amssymb,aps,prx,showpacs,floatfix,longbibliography]{revtex4-1}

\usepackage[dvips]{color,graphicx}
\usepackage[usenames,dvipsnames,svgnames,table]{xcolor}
\DeclareGraphicsExtensions{.eps, .jpg, .png}

\newcommand*{\ang}{\mbox{\normalfont\AA}}
\newcommand*{\eV}{\, \textrm{eV}}
\newcommand*{\meV}{\, \textrm{meV}}
\newcommand*{\eVmolecule}{\, \textrm{eV}/\textrm{H}_2\textrm{O}}
\newcommand*{\meVmolecule}{\, \textrm{meV}/\textrm{H}_2\textrm{O}}

\newcommand*{\meVdolecule}{\, \textrm{meV}/\textrm{D}_2\textrm{O}}

\begin{document}

\title{Anharmonic nuclear motion and the relative stability of
  hexagonal and cubic ice}

\author{Edgar A.\ Engel}
\affiliation{TCM Group, Cavendish Laboratory, University of Cambridge,
  J. J. Thomson Avenue, Cambridge CB3 0HE, United Kingdom}
\author{Bartomeu Monserrat}
\affiliation{TCM Group, Cavendish Laboratory, University of Cambridge,
  J. J. Thomson Avenue, Cambridge CB3 0HE, United Kingdom}
\author{Richard J.\ Needs} 
\affiliation{TCM Group, Cavendish Laboratory, University of Cambridge,
  J. J. Thomson Avenue, Cambridge CB3 0HE, United Kingdom}

\date{\today}

\begin{abstract}
We use extensive first-principles quantum mechanical 
calculations to show that, although the static lattice and harmonic 
vibrational energies are almost identical, the anharmonic vibrational 
energy of hexagonal ice is significantly lower than that of cubic ice.
This difference in anharmonicity is crucial, stabilising hexagonal 
ice compared with cubic ice by at least 1.4 $\meVmolecule$, in 
agreement with experimental estimates.
The difference in anharmonicity arises predominantly from molecular 
O-H bond stretching vibrational modes and is related to the different 
stacking of atomic layers.
\end{abstract}

\maketitle

\section{Introduction}
\begin{figure}
	\includegraphics[width=0.38\textwidth]{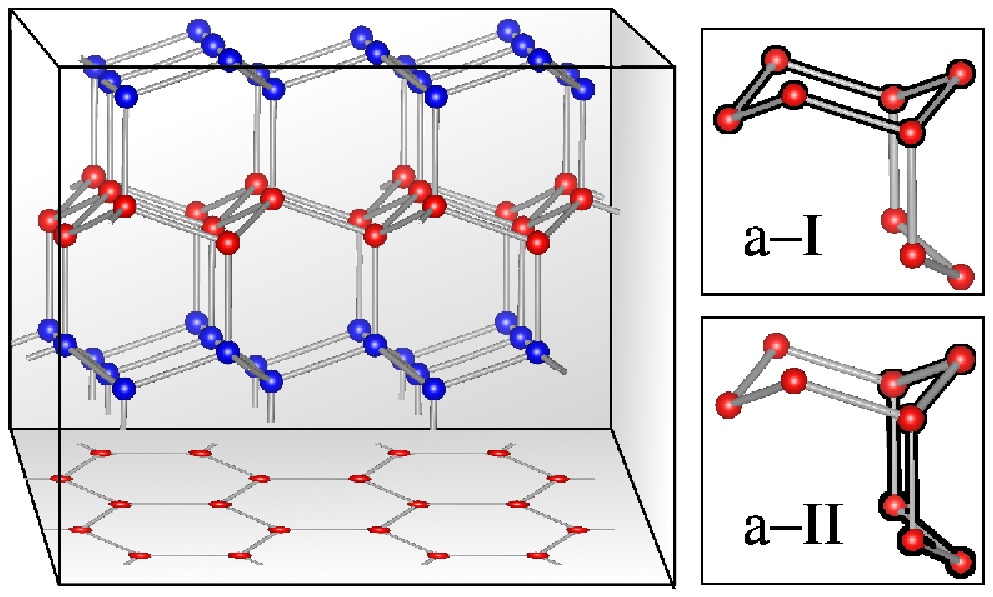}
	\vspace{0.25cm}\\
	\includegraphics[width=0.38\textwidth]{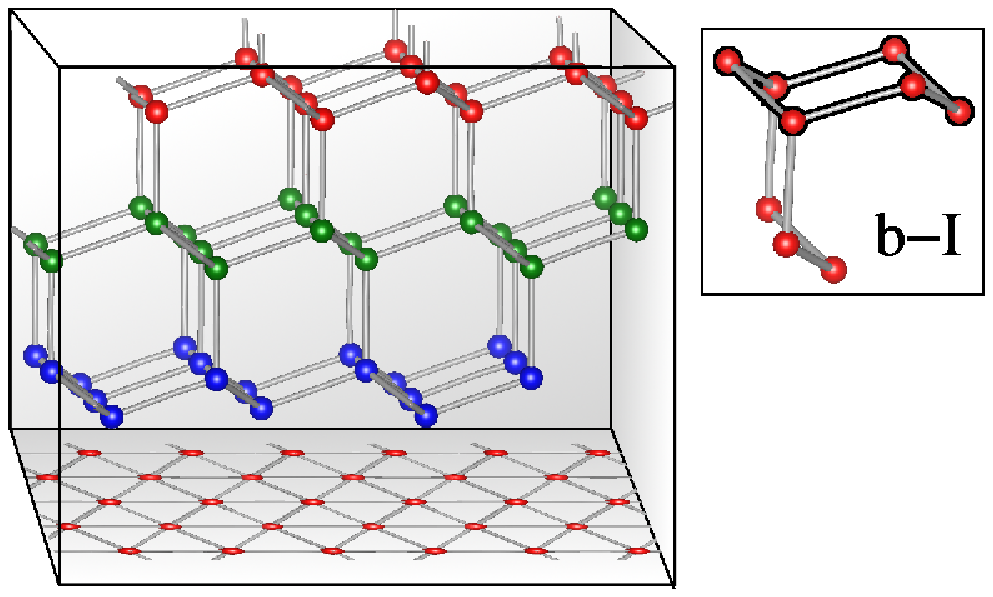}
	\caption{The oxygen sublattices in (a) hexagonal ice, Ih,
		and (b) cubic ice, Ic, consist of ABAB and ABC stacked 
		puckered layers (red, blue and green colours), respectively. 
		Projections of the periodic structures along the direction 
		orthogonal to the puckered layers are shown on the lower 
		faces of the diagrams.
		Ih is built of the chair- and boat-form hexamers shown in a-I 
		and a-II, respectively, while Ic is built exclusively from 
		chair-form hexamers as in b-I. The structures were visualised 
	using VESTA \cite{vesta}.}
	\label{fig:Stacking}
\end{figure}

Six-fold symmetric snow crystals are formed from hexagonal ice (Ih), 
which covers about 10\% of the Earth's surface and plays a prominent 
role in determining its climate \cite{bartels,baker,young}.
A cubic form of ice also occurs in nature, but is very rare 
\cite{murray_formation_ic,shilling_vapour_pressure_ic_clouds}.
The structures of pure hexagonal and cubic ice differ only in the 
stacking of layers of tetrahedrally coordinated water molecules (see 
Fig.\ \ref{fig:Stacking}). Yet, hexagonal ice is thermodynamically 
more stable than cubic ice, with experiments indicating the difference 
in stability to lie in the $\meVmolecule$ range \cite{carr,thuermer,
yamamuro_heat_capacity_ice, handa,mayer_cubic_ice_from_water,mcmillan,
sugisaki_cubic_ice_calorimetry,kuhs_stacking_disorder,
murray_kinetics_ice_freezing,steytler}.
There is a growing realisation that real ``cubic ice'' typically 
contains many stacking faults and is not pure cubic ice (Ic) 
as originally suggested by K\"{o}nig \cite{konig}. Stacking faulted 
ice is a highly complex material, whose nature and properties 
depend heavily on the free energy difference between pure Ih and Ic.

\begin{figure}[tbh]
	\centering
	\includegraphics[width=0.4\textwidth]{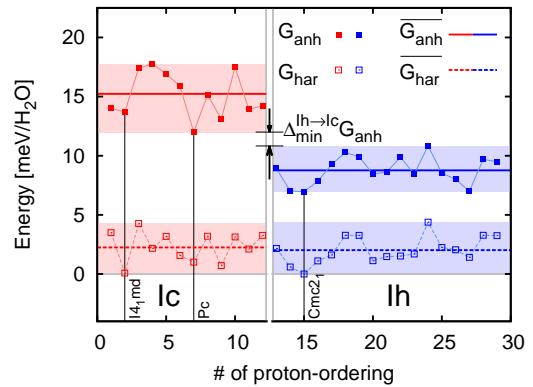}
	\caption{
		Harmonic ($G_{\textrm{har}}$, empty squares, lower part) and 
		anharmonic ($G_{\textrm{anh}}$, filled squares, upper part) energies 
		including zero-point nuclear motion for different Ih (blue) and Ic 
		(red) proton-orderings (Supplementary Table 1).
		The averages over proton-orderings, $\overline{G_{\textrm{har}}}$ 
		and $\overline{G_{\textrm{anh}}}$, are shown as thick horizontal dotted 
		and solid lines, respectively. All energies are measured with respect 
	to $G^{\textrm{XIh}}_{\textrm{har}}$ (Ih $Cmc2_{1}$, structure 15).}
	\label{fig:FreeEnergy}
\end{figure}

The very similar free energies of Ih and Ic have so far prevented 
state-of-the-art first-principles quantum mechanical calculations from 
explaining the stability of Ih. 
Both density functional theory (DFT) and diffusion quantum Monte Carlo 
studies have found that Ih and Ic are almost degenerate in energy 
when nuclear motion is neglected \cite{raza}.  Our calculations 
show that the harmonic zero-point vibrational energies of Ih and Ic are 
large, at roughly $700 \meVmolecule$, but they are almost identical. 
Consequently, when averaged over different proton-orderings, the two 
phases are almost degenerate when harmonic vibrations are included (see 
Fig.~\ref{fig:FreeEnergy}). 
However, the small mass of hydrogen gives rise to large amplitude 
vibrations and large anharmonic effects.

A substantial body of theoretical work exists on water and ice, based 
on force-field path-integral molecular dynamics (PIMD), first-principles 
classical molecular dynamics (MD), and first-principles 
vibrational calculations in the quasi-harmonic approximation. 
This work has led to significant successes in understanding the 
important role of quantum vibrations and anharmonicity for 
various phenomena observed experimentally. 
Examples include, but are not limited to, (a) the isotope effects, e.g., on 
the melting temperature, of Ih upon going from protiated to deuterated ice 
\cite{ramirez_PIMD_isotope_effect_melting,ramirez_PIMD_isotope_effects_ice,
pena_quantum_effects_ice}, (b) accurate O-H bond lengths and infrared O-H 
stretching frequencies in water \cite{habershon,zeidler_exp_quantum_isotope_effects,
ramirez_HDA_ice}, (c) reproduction of the anomalous thermal expansion, and 
the isotope effect on the volume in Ih \cite{pamuk} and (d) the heat capacity 
of water \cite{vega_heat_cap_signature_quantum_effects}.
Attempts to calculate the relative stability of Ih and Ic have either 
relied on empirical force-fields such as TIP4P \cite{tanaka,
quigley_thermodynamics_stacking_disorder} or have lacked an accurate 
description of anharmonicity \cite{raza,ramirez_ice_phase_diagram}. 
TIP4P has since been shown to produce incorrect proton-ordering energetics 
and an incorrect static lattice energy difference between Ih and Ic 
compared to highly accurate diffusion Monte Carlo methods \cite{singer_knight}. 
Moreover, no successful attempts to explain the origin of the greater 
stability of Ih have been made.
With our fully anharmonic, first-principles DFT study we show that the 
inclusion of accurate anharmonic quantum nuclear motion is decisive in 
stabilising Ih with respect to Ic, and relate the difference in stability 
to the different stacking of the atomic layers.



The water molecules in both Ih and Ic are tetrahedrally coordinated, 
each donating and accepting two hydrogen bonds and thus satisfying 
the ``Bernal-Fowler ice rules'' \cite{bernal_fowler}. The oxygen 
sublattices of Ih and Ic arise from an ABAB and an ABC stacking of 
puckered layers of oxygen atoms, respectively. Correspondingly, the 
basic building blocks of Ih are chair- and boat-form hexamers, while 
Ic is built exclusively from chair-form hexamers (Fig.\ \ref{fig:Stacking}).
Pure Ic has so far proven elusive. Experimentally synthesised ``Ic'' 
typically contains many stacking faults which strongly affect its 
physical and chemical properties \cite{malkin,kuhs_stacking_disorder,carr}. 
Ice containing cubic sequences interlaced with hexagonal sequences 
is commonly known as stacking-disordered ice (Isd).
Unlike Ih and Ic, which refer to a unique stacking arrangement of 
puckered layers, Isd refers to the infinite set of possible stacking 
sequences. This set smoothly connects Ih as one end member to Ic as 
the other.
Isd has trigonal $P3m1$ symmetry \cite{hansen1,hansen2,hallett}.
As of yet, it is unclear whether stacking disorder is kinetically or 
thermodynamically driven. 
A full understanding of Isd will require understanding the properties 
of Ih and Ic, including their relative stability. At the most basic 
level, the free energy difference between Ih and Ic is required 
to understand why Isd is found to anneal to Ih rather than Ic.

The fraction of cubic stackings of layers, or ``cubicity'', 
typically does not exceed around 60\% \cite{malkin} in experiments and both 
the fraction itself as well as the nature of the stacking arrangements 
depend heavily on the synthesis pathway 
\cite{malkin,kuhs_stacking_disorder,moore}.
Crucially, ice synthesised via both homogeneous and heterogeneous 
freezing of (supercooled) water has a random stacking of cubic and 
hexagonal layers \cite{malkin,moore}, which is consistent with a 
layer-by-layer growth mechanism. Heterogeneous freezing in particular 
is central to atmospheric and climate physics and, due to random 
stacking-disorder, depends vitally on the free energy difference between 
Ih and Ic. It is clear that Isd is an extremely important and highly 
complex material.

Molecular dynamics and Monte Carlo simulations using empirical ice 
potentials to model ice nucleation processes have successfully 
reproduced stacking-disorder (see \cite{quigley_thermodynamics_stacking_disorder,
moore,silva_stacking_disorder_energetics} and references therein). 
However, they struggle, amongst other issues, with the accuracy of 
the empirical potentials in describing the melting temperature and the 
relative stability of Ih and Ic \cite{pena_accuracy_of_empirical_potentials}. 

In the following we limit ourselves to the study of pure Ih and Ic.

\section{Computational Model}
Atomistic simulations of proton-disordered systems such as Ih and Ic 
require sets of explicit atomic positions and a calculation must be 
performed for each proton-ordering studied. 
The number of proton-ordered, energetically quasi-degenerate 
structures allowed by the ice rules increases exponentially with the 
size of the simulation cell.  
This leads to Pauling's residual configurational entropy \cite{pauling},
$S_{\textrm{config}}$, which has been confirmed experimentally 
\cite{tajima, jackson}.  For large systems with negligible surface
effects the associated configurational free energies of Ih and Ic, 
$\Delta G_{\textrm{config}} = - T S_{\textrm{config}}$, are almost 
identical \cite{nagle_lattice_statistics,ramirez_configurational_entropy}. 
We therefore neglect $\Delta G_{\textrm{config}}$ in the following.

To gain an understanding of the effects of proton-ordering on the
vibrational properties of ice, we consider 16 distinct proton-ordered
eight-molecule Ih configurations as constructed by Hirsch and Ojam\"ae
\cite{hirsch}, and $11$ distinct proton-ordered eight-molecule Ic
configurations \cite{raza}.  We also con\-si\-der the ``con\-ventio\-nal'' 
he\-xa\-go\-nal, 12-mo\-le\-cule $P6_{3}cm$ Ih and quasi-cubic, eight-molecule 
$P4_{3}$ Ic structures (numbers $13$ and $1$ in Fig.\ \ref{fig:FreeEnergy}, 
respectively).  Details of the proton-ordered structures and their numbering 
are provided in Supplementary Table 1.

We performed electronic structure calculations using plane-wave 
pseudopotential DFT as implemented in the {\sc castep} code 
\cite{clarkSPHPRP05} (version 7.02). We employed the Perdew-Burke-Ernzerhof 
(PBE) \cite{perdew_1996_PBE} generalised gradient approximation to the 
exchange-correlation functional, and on-the-fly generated ultrasoft 
pseudopotentials \cite{Vanderbilt90} with core radii of $0.7\ \ang$ and 
$0.8\ \ang$ for the hydrogen and oxygen atoms, respectively. 
Supplementary Section V describes results obtained with other density 
functionals.  We used a plane-wave energy cut-off of $1600 \eV$ and 
Monkhorst-Pack reciprocal space grids of spacing less than 
$2\pi \times 0.04\ \ang^{-1}$ for all total energy calculations and 
geometry optimisations.  The resulting energy differences between 
frozen-phonon configurations are converged to below $10^{-4}\eVmolecule$, 
the atomic positions are converged to within $10^{-5}\ang$, and the 
residual forces to within $10^{-4}\eV/\ang$.

We obtained harmonic vibrational free energies from the harmonic 
frequencies calculated using the $\mathbf{k}$-space Fourier 
interpolated dynamical matrix.  The latter was obtained by Fourier 
transforming the real-space matrix of force constants constructed 
using a finite displacement method.  
Anharmonic vibrational free energies were calculated using the
method described in \cite{anharmonic_method}, which has so far been 
successfully applied to high-pressure systems \cite{monserrat,azadi}. 
As in \cite{anharmonic_method}, we describe the 3${\cal N}$-dimensional BO energy 
surface (where ${\cal N}$ is the number of atoms in the simulation cell) 
by mapping 1D subspaces along the harmonic normal mode axes up to 
large amplitudes of four times the harmonic root-mean-square (RMS) displacements, 
where anharmonicity is important. We then reconstruct the 3${\cal N}$-dimensional 
BO surface from the 1D subspaces. The resultant representation of the BO energy 
surface is an approximation to the true 3${\cal N}$-dimensional BO energy surface.
This approximation only weakly affects the free energy difference 
between Ih and Ic (see Supplementary Section VI). 
The 1D energy surfaces were fitted using cubic splines. 
The anharmonic vibrational Schr\"odinger equation was solved within a 
vibrational self-consistent field (VSCF) framework. 
The vibrational wave function was expanded in a basis of simple 
harmonic oscillator eigenstates, and the inclusion of $25$ states for each 
vibrational degree of freedom was found sufficient to obtain converged results.
  
\section{Results}
Our calculations show that the static lattice energies of Ih
and Ic, $E_{\textrm{static}}$, vary by up to $5 \meVmolecule$ with 
proton-ordering. This agrees with Refs.~\cite{raza,
hirsch,lekner_hydrogen_ordering_energetics} and, more importantly, 
with Ref.~\cite{singer}, which evaluates DFT static lattice 
energies for 16 8-molecule orthorhombic, 14 12-molecule hexagonal 
and 63 48-molecule orthorhombic Ih proton-orderings.
This strongly suggests that our sets of proton-orderings provide a 
good representation of the distribution of energies in disordered 
ice.

We also find that the harmonic vibrational contributions to the free 
energies of different proton-orderings, $\Delta G_{\textrm{har}}$, vary 
by up to $2\meVmolecule$.  We have employed the VSCF method described 
above to calculate the anharmonic contribution to the vibrational 
free energy, $\Delta G_{\textrm{anh}}$, finding a variation between 
proton-orderings of up to $5 \meVmolecule$

The free energies of Ih and Ic at the harmonic vibrational level,
$G_{\textrm{har}}=E_{\textrm{static}}+\Delta G_{\textrm{har}}$, 
averaged over the proton-orderings, are virtually indistinguishable:
\begin{equation*}
	\Delta_{\mathrm{av}}^{\mathrm{Ic} \rightarrow \mathrm{Ih}}G_{\mathrm{har}} 
	\equiv \overline{G_{\textrm{har}}^{\textrm{Ic}}}
	- \overline{G_{\textrm{har}}^{\textrm{Ih}}} 
	= 0.2 \pm 2.4 \meVmolecule,
\end{equation*}
where the quoted errors are the RMS variations across 
different proton-orderings. The anharmonic energies of the Ih 
configurations, $\Delta G_{\textrm{anh}}$, on the other hand, though 
also positive, are systematically lower than those of the Ic 
configurations, so that the total free energy, $G_{\textrm{anh}} = 
E_{\textrm{static}} + \Delta G_{\textrm{har}} + \Delta G_{\textrm{anh}}$, 
averaged over the different proton-orderings, is significantly lower 
for Ih than for Ic:
\begin{equation*}
	\Delta_{\mathrm{av}}^{\mathrm{Ic} \rightarrow \mathrm{Ih}}G_{\mathrm{anh}} 
	\equiv \overline{G_{\textrm{anh}}^{\textrm{Ic}}}
	- \overline{G_{\textrm{anh}}^{\textrm{Ih}}}
	= 6.5 \pm 3.1\meVmolecule .
\end{equation*}
The values obtained for
$\Delta_{\mathrm{av}}^{\mathrm{Ic} \rightarrow \mathrm{Ih}}G_{\mathrm{har}}$
and
$\Delta_{\mathrm{av}}^{\mathrm{Ic} \rightarrow \mathrm{Ih}}G_{\mathrm{anh}}$
depend significantly on the method of averaging. For example, using 
a Boltzmann distribution for the free energy of the proton-orderings 
leads to values of 
$\Delta_{\mathrm{Boltzmann}}^{\mathrm{Ic} \rightarrow \mathrm{Ih}}G_{\mathrm{har}}
\approx 5.5\meVmolecule$ and $6.1 \meVmolecule$ at 10 and 100 K,
respectively.

\begin{figure}
	\centering
	\includegraphics[width=0.4\textwidth]{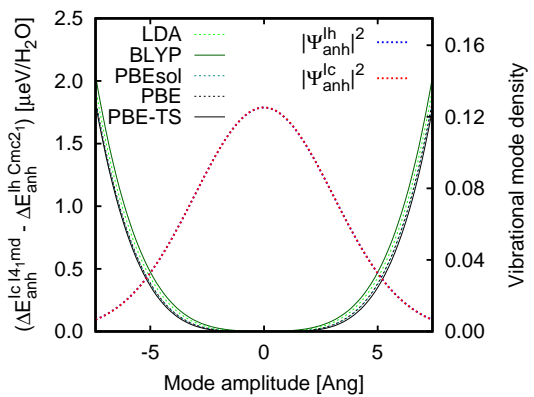}
	\caption{Difference in the anharmonicity of the BO energy surface, 
		$\Delta E_{\textrm{anh}}$, of Ih $Cmc2_1$ and Ic $I4_1md$ for 
		various functionals (thin solid and dotted lines). 
		The dominant contribution to the anharmonicity is quartic.
		The difference between the results from different functionals 
		is so small it can barely be distinguished.
		The different widths of the vibrational densities, 
		$\vert\Psi_{\textrm{anh}}\vert^2$, of Ih and Ic (blue and red 
		thick dotted lines, respectively) is indistinguishable on the 
		scale of the figure.}
	\label{fig:XC_Anh}
\end{figure}

It is noteworthy that, given cell volumes that are reasonably close to
experiment, the differences in $E_{\textrm{static}}$, 
$\Delta G_{\textrm{har}}$ and $\Delta G_{\textrm{anh}}$ between Ih and Ic
depend only weakly on the details of the DFT calculations and in 
particular on the choice of exchange-correlation functional (see 
Fig.\ \ref{fig:XC_Anh} and Supplementary Section V).

The most stable proton-ordered configurations of Ih and Ic, referred
to as XIh and XIc, display free energy differences very similar to
$\Delta_{\mathrm{av}}^{\mathrm{Ic} \rightarrow \mathrm{Ih}}G_{\mathrm{har}}$
and
$\Delta_{\mathrm{av}}^{\mathrm{Ic} \rightarrow \mathrm{Ih}}G_{\mathrm{anh}}$.
However, the inclusion of anharmonic vibrational energies chan\-ges the 
relative stability of the different proton-orderings. XIh is experimentally 
and theoretically known to have space group $Cmc2_1$
\cite{experiment_Cmc21,hirsch} (structure $15$ in Fig.\ \ref{fig:FreeEnergy}), 
a result confirmed by our calculations. 
A structure of space group $I4_1md$ has been proposed \cite{raza} (structure 
$2$) for XIc on theoretical grounds.  Our calculations support this proposal 
at the harmonic vibrational level, but the inclusion of anharmonic 
contributions suggests that Ic $Pc$, $Pca2_1$ and  $Pna2_1$ (structures $7$, 
$9$ and $12$) may also be strong candidates for XIc. 
Ref.\ \cite{geiger} reports experimental evidence for Ic $I4_1md$ and 
$Pna2_1$ in partially proton-ordered Ic via Fourier transform infrared 
spectroscopy.  This lends significant support to our result that 
anharmonicity provides the decisive contribution to the energy differences 
between proton-orderings, since $Pna2_1$ has a high free energy at the 
static lattice and harmonic vibrational levels and only becomes a 
low-free-energy structure when anharmonicity is taken into account.


At typical experimental temperatures of below 100 K, the
proton-ordering is largely frozen in. Consequently one cannot expect
to measure a change in free energy corresponding to a transition from
Ic to Ih in thermal equilibrium, but rather (assuming the Ic sample is
annealed at low temperatures and consists mostly of XIc) a change in
free energy corresponding to transitions from XIc to a proton-ordering
of Ih, which is likely to be smaller than
$\Delta_{\mathrm{av}}^{\mathrm{Ic} \rightarrow
  \mathrm{Ih}}G_{\mathrm{anh}}$.  As indicated in Fig.\
\ref{fig:FreeEnergy}, we evaluate a lower bound on the
free energy difference as
\begin{equation*}
   \begin{split}
		\Delta_{\mathrm{min}}^{\mathrm{Ic} \rightarrow \mathrm{Ih}}G_{\mathrm{anh}} 
      & \equiv 
		\mathrm{min}\left\{G_{\mathrm{anh}}^{\mathrm{Ic}}\right\} -
		\mathrm{max}\left\{G_{\mathrm{anh}}^{\mathrm{Ih}}\right\} \\
      & = 1.4 \pm 0.3 \meVmolecule .
   \end{split}
\end{equation*}
This lower bound is consistent with, but on the high side of,
experimentally measured free energy differences of $0.3 - 1.6
\meVmolecule$ \cite{thuermer, yamamuro_heat_capacity_ice, handa,
  mayer_cubic_ice_from_water, mcmillan,
  sugisaki_cubic_ice_calorimetry, kuhs_stacking_disorder,
  murray_kinetics_ice_freezing, steytler}.  Notably, the experimental
value is rather uncertain, mainly because the free energy difference
is very small, and because the Ic samples are typically not fully 
characterised in terms of stacking faults or proton ordering.

\begin{figure}[tbh]
   \centering
   \includegraphics[width=0.4\textwidth]{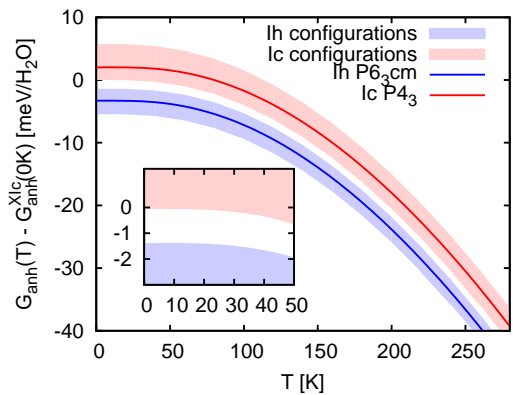}
   \caption{$G_{\textrm{anh}}$ for H$_2$O as a function of temperature. 
		The solid blue and red lines indicate results for Ih $P6_{3}cm$
		and Ic $P4_{3}$ simulation cells containing 96 and 64 molecules,
      respectively.
      The light blue and light red areas indicate the spread in free
   energies arising from the different Ih and Ic proton-orderings.}
   \label{fig:FreeEnergy_TempDep}
\end{figure}
As shown in Fig.\ \ref{fig:FreeEnergy_TempDep}, this energy 
difference remains roughly constant from zero temperature up 
to 273 K and thus stabilises Ih over a wide range of temperatures.

\begin{figure}
   \centering
      \includegraphics[width=0.4\textwidth]{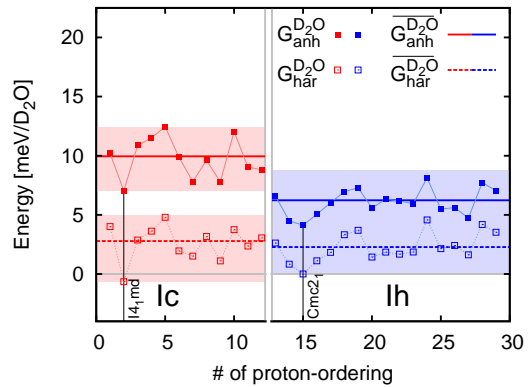}
   \caption{
		Zero temperature $G^{\textrm{D}_2\textrm{O}}_{\textrm{har}}$ (empty 
		squares, lower part) and $G^{\textrm{D}_2\textrm{O}}_{\textrm{anh}}$ 
		(filled squares, upper part) for different deuterated Ih (blue) 
		and Ic (red) proton-orderings. The averages over proton-orderings, 
		$\overline{G^{\textrm{D}_2\textrm{O}}_{\textrm{har}}}$ and 
		$\overline{G^{\textrm{D}_2\textrm{O}}_{\textrm{anh}}}$, are shown as thick 
		dotted and solid lines, respectively. All energies are measured 
	with respect to $G^{\textrm{D}_2\textrm{O}}_{\textrm{har}}$ of Ih $Cmc2_{1}$.}
	\label{fig:FreeEnergy_HeavyIce}
\end{figure}
In D$_2$O, the heavier mass of deuterium firstly leads to reduced 
vibrational frequencies of the harmonic vibrational modes and 
thus reduced vibrational energies. For XIh D$_2$O, 
$\Delta G_{\textrm{har}} = 507.67 \meVdolecule$ compared to 
$692.34 \meVmolecule$ for XIh H$_2$O.
Secondly, D$_2$O has smaller vibrational amplitudes and consequently 
smaller anharmonic vibrational energies than H$_2$O, 
resulting in a smaller difference between the free energies of Ih 
and Ic,
\begin{equation*}
	\Delta_{\mathrm{av}}^{\mathrm{Ic} \rightarrow \mathrm{Ih}}
	G_{\mathrm{anh}}^{\textrm{D$_2$O}} \equiv
	\overline{G_{\textrm{anh}}^{\textrm{Ic-D$_2$O}}}
	- \overline{G_{\textrm{anh}}^{\textrm{Ih-D$_2$O}}}
	= 3.7 \pm 2.9\meV/\textrm{D}_2\textrm{O} ,
\end{equation*}
as shown in Fig.\ \ref{fig:FreeEnergy_HeavyIce}. This suggests 
that it could be easier to synthesise deuterated than protiated 
cubic ice.
Moreover, the smaller anharmonic energy calculated for D$_2$O 
compared with H$_2$O implies that the most likely candidate for 
the ground state of cubic D$_2$O has $I4_1md$ symmetry (structure $2$) 
and is thus different from the ground state of H$_2$O. The predicted 
ground state of hexagonal D$_2$O is $Cmc2_1$, as for H$_2$O.

\begin{figure}[bth]
	\centering
	\includegraphics[width=0.4\textwidth]{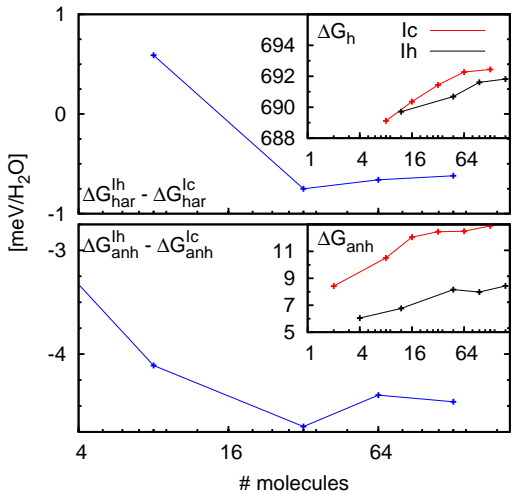}
	\caption{The harmonic free energy, $\Delta G_{\textrm{har}}$, and anharmonic
		free energy, $\Delta G_{\textrm{anh}}$, are shown as functions of the
	simulation cell size in the upper and lower panels, respectively.}
	\label{fig:SCConvergence}
\end{figure}

We have investigated the convergence of $\Delta G_{\textrm{har}}$ 
and $\Delta G_{\textrm{anh}}$ with respect to the size of the simulation 
cell using cells containing up to 192 molecules for Ih $P6_3cm$ and 128 
molecules for Ic $P4_3$ as shown in Fig.\ \ref{fig:SCConvergence}.

The vibrational energies of the different proton-orderings were
calculated using 64-molecule cells for the harmonic vibrational energy
and eight-molecule cells for the anharmonic energies.  The latter is 
justified by the short-range nature of anharmonicity, evidence for which 
is described in section \ref{Discussion}.  
Calculations using cells with up to $192$ molecules for the $P6_{3}cm$ Ih 
structure and $128$ molecules for the $P4_{3}$ Ic structure indicate that 
the difference between $\Delta G_{\textrm{har}}$ for Ih and Ic is converged 
to within $0.1 \meVmolecule$ using 64-molecule simulations cells. The
anharmonic corrections converge analogously, and the difference 
between $\Delta G_{\textrm{anh}}$ for the two structures is
converged to within $0.2 \meVmolecule$ using 64-molecule simulations cells. 
More details may be found in Supplementary Section II.

\section{Discussion}
\label{Discussion}
\begin{figure}
   \centering
   (a) Ih vibrational bandstructure and DoS.\\
   \vspace{0.25cm}
   \includegraphics[width=0.4\textwidth]{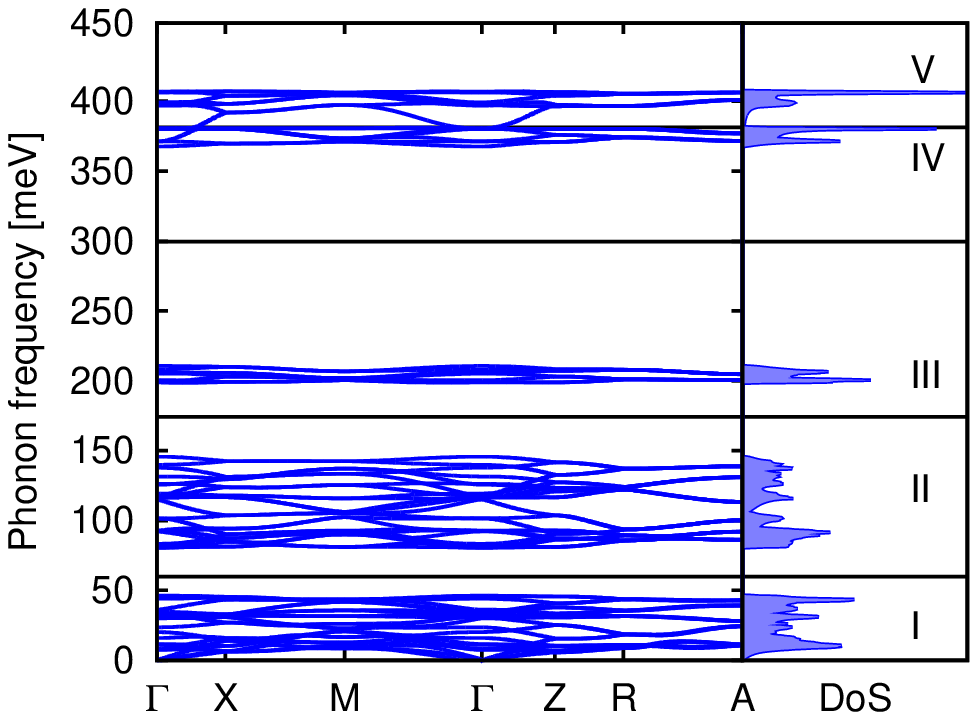} \\
   \vspace{0.25cm}
   (b) Ic vibrational bandstructure and DoS.\\
   \vspace{0.25cm}
   \includegraphics[width=0.4\textwidth]{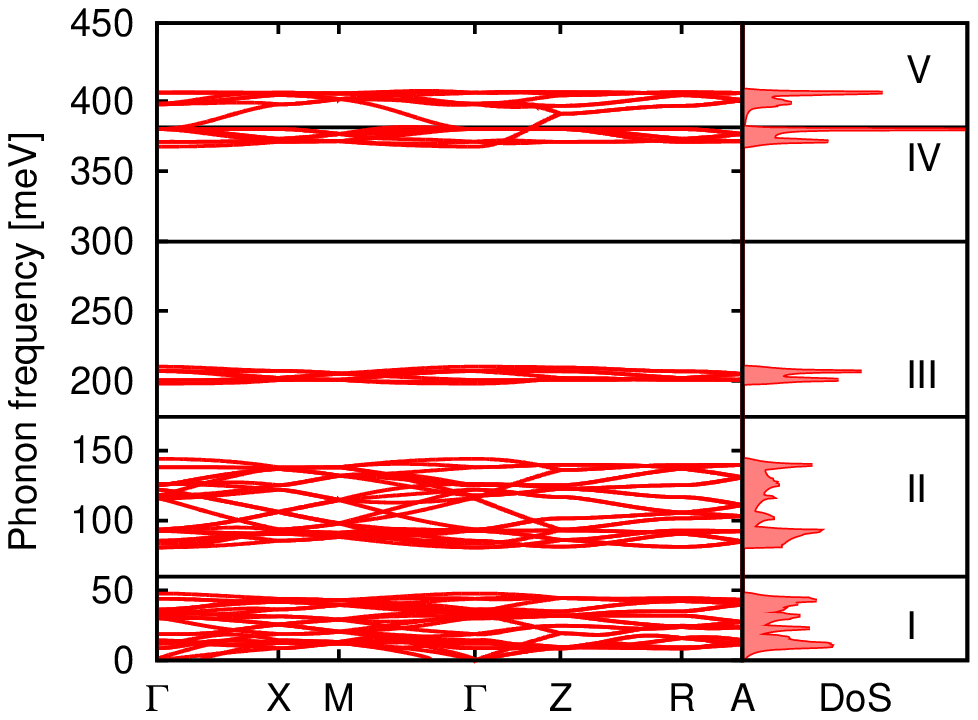} \\
	\caption{Vibrational bandstructure and DoS of Ih (a) and Ic (b) 
		split into crystal modes (I), libration modes (II), molecular 
		O-H bending modes (III) and molecular O-H anti-symmetric 
	and symmetric stretching modes (IV and V, respectively).}
	\label{fig:VibrBandStruct}
\end{figure}

\begin{figure}
\centering
\includegraphics[width=0.4\textwidth]{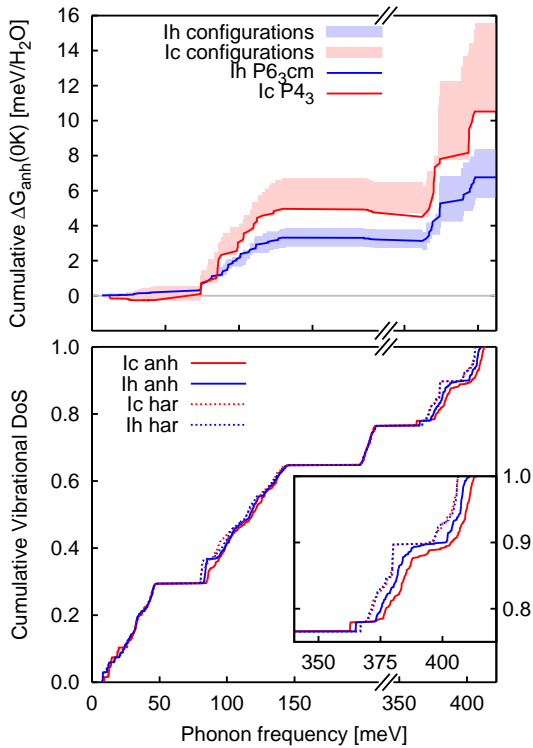}
\caption{Cumulative anharmonic energies and vibrational DoS of 
	H$_2$O averaged over proton-orderings as a function of frequency. 
	The upper panel shows the anharmonic energies for Ih $Cmc2_{1}$ 
	and Ic $P4_{3}$ as solid blue and red lines and the spread in 
	free energies in Ih and Ic due to different proton-orderings as 
	light red and light blue areas.
	The lower panel shows that the harmonic DoS of Ih and Ic 
	(dotted lines) are practically identical. The larger increase in 
	vibrational energy due to anharmonicity in Ic than in Ih is 
	reflected in the shifts of their anharmonic DoS (continuous lines) 
with respect to the harmonic DoS.}
\label{fig:PhononDoSAndAnharmCorrVsFreq}
\end{figure}

The origin of the difference in the anharmonicities of Ih and Ic can be 
traced back to the libration ($80 - 140 \meV$) and, predominantly, the 
molecular symmetric and anti-symmetric O-H bond stretching modes 
($365 - 380 \meV$ and $395 - 410 \meV$, respectively) indicated in 
Fig.\ \ref{fig:VibrBandStruct}.
The vibrational density of states (DoS) and the distribution of 
anharmonic corrections over the vibrational frequencies show 
(Fig.\ \ref{fig:PhononDoSAndAnharmCorrVsFreq}) that the dominant 
anharmonic contributions arise from the O-H bond stretching modes, 
which correspond to large amplitude displacements of the hydrogen 
atoms relative to their neighbouring (essentially stationary) 
oxygen atoms. 
The comparatively small role of the oxygen atoms is confirmed by 
studying Ih and Ic analogues with fixed oxygen positions, which 
recover the value of $\Delta G_{\textrm{anh}}$ observed for real Ih 
and Ic to within 5\% (see Supplementary Table III).
The O-H bond stretching modes contribute $> 2/3$ of the difference 
in anharmonicity between Ih and Ic, when averaged over proton-orderings.
Note that the energy difference between Ih and Ic shown in Fig.\ 
\ref{fig:PhononDoSAndAnharmCorrVsFreq} increases at high frequencies.

Variations in vibrational frequencies with cubicity have recently been 
identified in infrared absorption experiments \cite{carr}. 
Carr \textit{et al.}\ observed that the O-H stretching modes were shifted 
to higher vibrational frequencies with increasing cubicity of their Isd 
samples. They also observed an increasing broadening of the absorption 
peak. According to Carr \textit{et al.}, both trends are thought to be 
associated with the stacking disorder, which peaks at a cubicity of 50\%. 
For samples with cubicities of 50\% Carr \textit{et al.}\ observed 
shifts of $28\pm2$ cm$^{-1}$ and $13\pm2$ cm$^{-1}$ for protiated and 
deuterated Isd, respectively. Our calculations reproduce the widening 
of the O-H stretching peak in the vibrational DoS and produce shifts of 
$70\pm5$ cm$^{-1}$ and $30\pm5$ cm$^{-1}$ for protiated and deuterated 
Isd, respectively. I.e., at about double the cubicity of Carr \textit{et al.}, 
we calculate shifts that are about twice as large as those of Carr 
\textit{et al.}\ Thus our values are consistent with the results of Carr 
\textit{et al.}\ if the shift scales roughly linearly with the degree 
of cubicity rather than the amount of stacking disorder, and is largest 
in pure Ic.
In accurately reproducing the ratio between the blueshifts of the molecular 
stretching frequencies for the protiated and deuterated phases, our results 
also provide a good account of the effects of isotopic substitution. 

The role of the high energy modes can be further illuminated by 
considering the H-H radial distribution functions (Fig.\ \ref{fig:RDFs} 
(b)) of Ih and Ic and RMS displacements of the protons (see Supplementary 
Table IV). The latter show that the harmonic vibrational amplitudes in 
Ic are about 1\% smaller than in Ih.
\begin{figure}
	\centering
	(a) O-H RDFs, $g_{\textrm{O-H}}(r)$.\phantom{blablablablablablablablabla}\\
	\vspace{0.25cm}
	\includegraphics[width=0.4\textwidth]{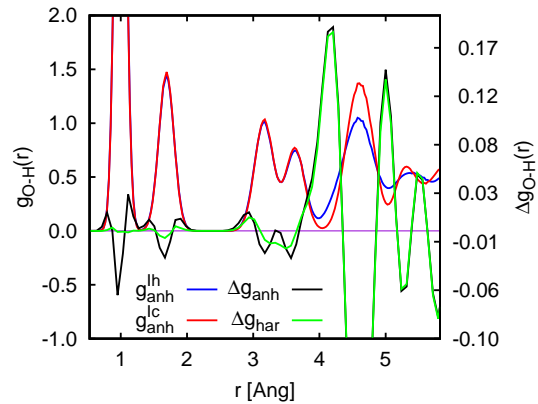} \\
   \vspace{0.25cm}
	(b) H-H RDFs, $g_{\textrm{H-H}}(r)$.\phantom{blablablablablablablablabla}\\
   \vspace{0.25cm}
	\includegraphics[width=0.4\textwidth]{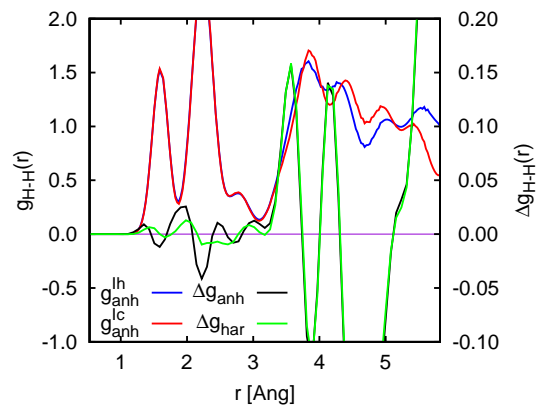} \\
	\caption{The RDFs for Ih $P6_3cm$ (blue) and Ic $P4_3$ (red) are sampled
		from the anharmonic nuclear wavefunctions. For radii $< 3\,$\AA\ the 
		difference between Ih and Ic at the harmonic level (green), 
		$\Delta g_{\textrm{har}} \equiv g^{\textrm{Ih}}_{\textrm{har}} 
		- g^{\textrm{Ic}}_{\textrm{har}}$, are minimal. 
		The difference at the anharmonic level (black), 
		$\Delta g_{\textrm{anh}} \equiv g^{\textrm{Ih}}_{\textrm{anh}} 
		- g^{\textrm{Ic}}_{\textrm{anh}}$, is small, but non-negligible.
		The features in $\Delta g_{\textrm{har}}$ and $\Delta g_{\textrm{anh}}$
		for radii beyond around $3\,$\AA\ originate predominantly from
	differences in the static structures of Ih and Ic.}
	\label{fig:RDFs}
\end{figure}
Yet, at the harmonic level the H-H radial distribution functions 
of Ih and Ic are essentially identical for distances less than about 
$3\,$\AA. Beyond $3\,$\AA\ the static structures of the two phases differ. 
Unlike the RMS displacements of the protons, the RDFs measure a 
two-particle quantity that describes correlated motions of pairs of 
protons. 
At the harmonic level the protons in Ic move less with respect to their 
equilibrium positions than in Ih, but they move by just as much with 
respect to each other.
We note that the O-H and H-H RDFs for Ih shown in Figs.\ \ref{fig:RDFs} 
(a) and (b) agree well with, e.g., the experimental RDFs in 
\cite{soper}.

\begin{figure}
   \centering
   (a) Typical chair-form configuration found in Ih and Ic.\\
   \vspace{0.25cm}
   \includegraphics[width=0.35\textwidth]{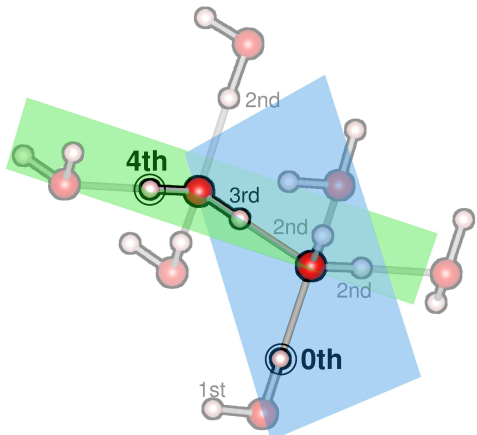}\\
   \vspace{0.25cm}
   (b) Typical boat-form configuration found only in Ih.\\
   \vspace{0.25cm}
   \includegraphics[width=0.35\textwidth]{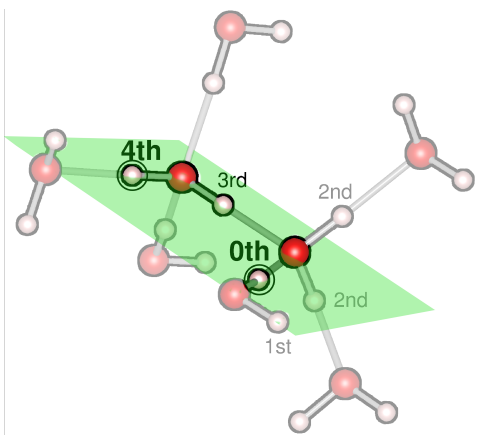}\\
   \caption{The (arbitrarily picked) central proton is labelled as
      ``0\textsuperscript{th}''. The nearest- to fourth-nearest-neighbours are
      labelled ``1\textsuperscript{st}'' through ``4\textsuperscript{th}''.
      From the point of view of the central proton, the 
		fourth-nearest-neighbour protons first lead to a distinction 
		between chair- and boat-form hexamers.
      The fourth-nearest-neighbour protons are found at distances
      from the central proton ranging from around 3.5 to 5.0 \AA.
		The green and blue planes are spanned by the relevant O-O axes 
		connecting the  fourth-nearest-neighbour protons.}
   \label{fig:HH_Neighbours}
\end{figure}

While the protons in both phases feel the same local environment, 
differences occur starting with the fourth-nearest-neighbour protons
(see Fig.\ \ref{fig:HH_Neighbours}).
Also, for systems as small as 8 to 12 
molecules $\Delta G_{\textrm{anh}}$ is already about 3/4 of the 
converged value (Supplementary Fig.\ 1). This system size limits the 
wavelength of the vibrational modes responsible for the difference 
in anharmonic energies to roughly the same distance as the separation 
of fourth-nearest-neighbour protons.
Together these observations indicate that the influence of more 
distant nuclei is small.
\begin{figure*}
   \centering
   \includegraphics[width=0.8\textwidth]{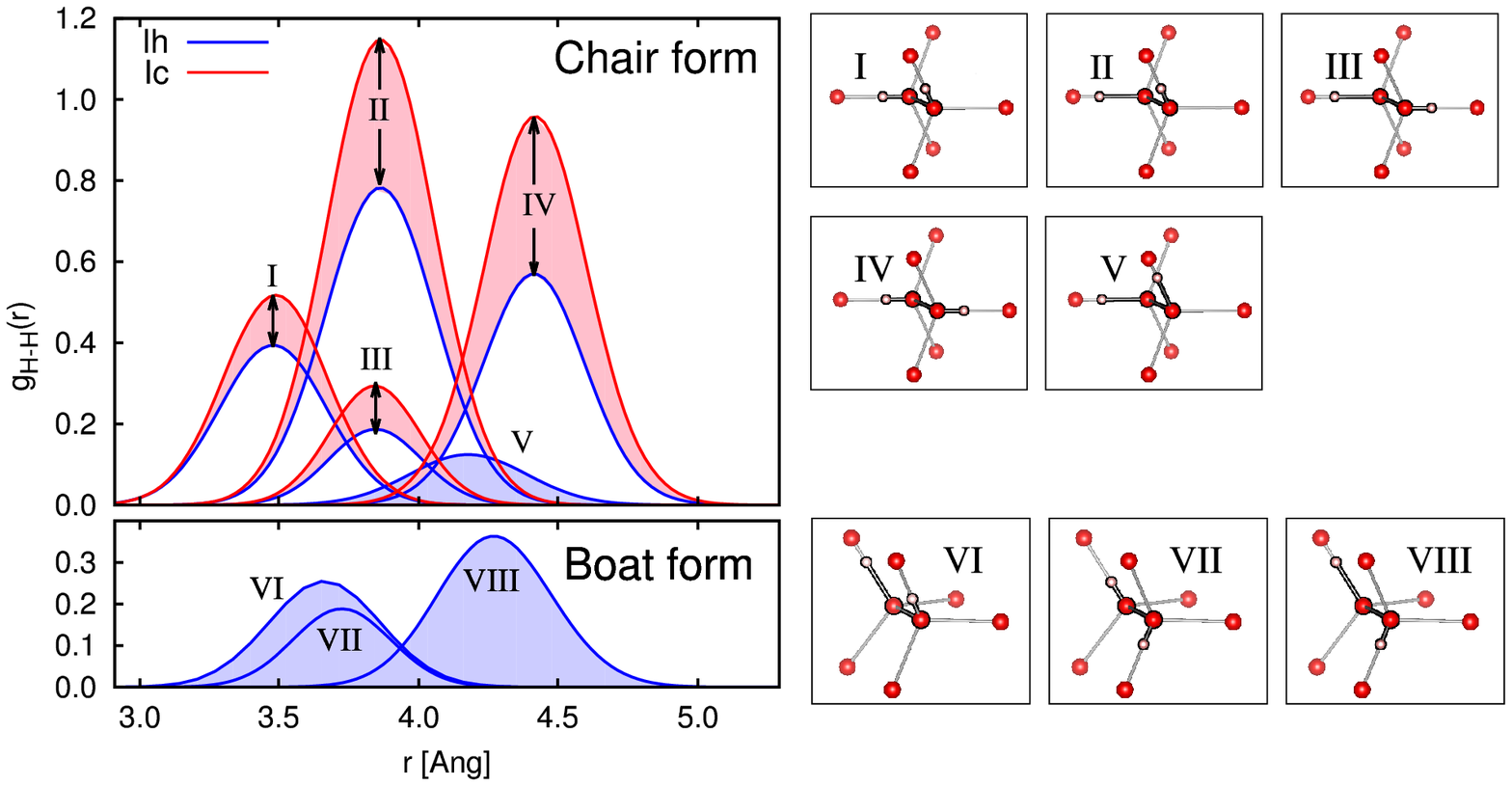}
	\caption{Anharmonic H-H RDF decomposed into contributions from 
			different bonding configurations of fourth-nearest-neighbour pairs 
		of proton. The upper part shows the components corresponding to 
		the chair-form hexamer configurations, I -- V, found in both Ih 
		and Ic. The lower part shows the components corresponding to 
		the boat-form hexamer configurations, VI -- VIII, which are exclusive 
		to Ih. In configurations VI -- VIII the hydrogen bridge bonds associated 
		with the pairs of protons (shown as small white spheres) are at a larger 
		angle to each other than in I -- V.}
	\label{fig:RDFs_Components}
\end{figure*}
Allowing for both chair- and boat-form hexamers, there are 12 distinct 
arrangements of fourth-nearest-neighbour pairs of protons. Out of 
these 12 arrangements only 8 are realised in the proton-orderings we 
have considered. These are shown in Fig.\ \ref{fig:RDFs_Components}.
Three of these arrangements (numbers VI -- VIII) are associated with 
boat-form hexamers of H$_2$O molecules, which only exist in Ih, 
and 5 (numbers I -- V) are associated with the chair-form hexamers of 
H$_2$O molecules found in Ih and Ic.
The RDFs in Fig.\ \ref{fig:RDFs_Components} show that, on going 
from Ic to Ih, arrangements II and IV are depopulated. 
For arrangements II and IV the displacement of the first proton of 
the pair from its equilibrium position along its hydrogen bridge bond 
leads to a large displacement relative to the second proton.
Conversely, arrangements VI -- VIII, for which the same displacement of 
the first proton leads to a far smaller displacement relative to the 
second, are populated. 
This explains why on average the protons in Ic move less with respect to 
their equilibrium positions than in Ih, while moving by just as much with 
respect to each other, resulting in the same $G_{\textrm{har}}$ as in Ih.

Going beyond the harmonic approximation, anharmonicity reduces the RMS 
vibrational amplitudes in Ih and Ic by around 1\% and 2.5\%, 
respectively, localising the nuclear wavefunction more in ice 
Ic than Ih.
On the level of the collective vibrational modes, the localisation is 
typically driven by a strong quartic contribution to the respective 
BO energy surfaces (see Supplementary Fig. 6).
Moreover, at the anharmonic level the RDFs show that the protons in ice 
Ic move less with respect to each other than in Ih, instead of just 
moving less with respect to their equilibrium positions, as they do at 
the harmonic level.
This difference in the relative motion of pairs of protons is the origin 
of the difference in $\Delta G_{\textrm{anh}}$ between Ih and Ic.
The larger effect of anharmonicity in Ic is again due to the stronger 
geometric ``coupling'' between pairs of protons.
\\
\section{Perspective}
Ih and Ic are important in various branches of science.  
Examples include climate modelling and the simulation of ice
nucleation and formation, where cubic ice plays an important role and
for which $\Delta G_{\textrm{anh}}$ is an essential input parameter.
As an example of the relevance in biological sciences, the benign shape 
of cubic ice crystals is of potential interest for cryopreservation 
\cite{mehl}.  
Here we have demonstrated that accounting for anharmonic nuclear 
vibrations is central to understanding and correctly predicting the 
free energy difference between Ih and Ic.  However, the importance of 
anharmonic vibrations in hydrogen-bonded systems reaches far beyond ice.  
An accurate treatment of anharmonicity is likely to be crucial in 
correctly describing the energy differences between very similar 
polymorphs of hydrogen-bonded molecular crystals which are important 
in, e.g., pharmaceutical materials science.

Calculating anharmonic vibrational energies in solids is a challenging 
computational task, which has only recently been successfully achieved 
using first-principles quantum mechanical methods. Anharmonic effects 
are particularly important for light elements, such as the hydrogen 
atoms in H$_2$O. Anharmonic vibrations are also expected to be important at 
the surfaces of ice, and when impurities or other defects are present.

\vspace{0.25cm}
We acknowledge financial support from the Engineering and Physical 
Sciences Research Council of the UK [EP/J017639/1]. 
B.\ M.\ also acknowledges Robinson College, Cambridge, and the Cambridge 
Philosophical Society for a Henslow Research Fellowship.
The calculations were performed on the Cambridge High Performance Computing 
Service facility and the HECToR and Archer facilities of the UK's national 
high-performance computing service (for which access was obtained via the 
UKCP consortium [EP/K013564/1]).

\bibliography{ice_Ih_Ic_final}


\end{document}